\documentclass[a4paper,10pt]{article}

\usepackage[dvips]{graphicx,psfrag}
\usepackage{amssymb}
\usepackage{amsmath}

\makeatletter

 \@addtoreset{equation}{section}
\makeatother
\newcommand{\beq}{\begin{eqnarray}}
\newcommand{\eeq}{\end{eqnarray}}
\def\tr{\mathop{\mathrm{tr}}\nolimits}

\begin{document}


\vspace*{-1cm}
\begin{flushright}
{\tt hep-th/0511300}\\
December 2005
\end{flushright}

\begin{center}
{\LARGE\bf
Wilsonian renormalization group approach to the lower dimensional nonlinear sigma models
}

\setcounter{footnote}{0}

{\renewcommand{\thefootnote}{\fnsymbol{footnote}}
{\large\bf Kiyoshi Higashijima $^1$\footnote{
     E-mail: {\tt higashij@phys.sci.osaka-u.ac.jp}}and
 Etsuko Itou$^2$\footnote{
     E-mail: {\tt itou@hep.s.kanazawa-u.ac.jp}} 

}}


{\sl
$^1$Department of Physics,
Graduate School of Science, Osaka University,\\ 
Toyonaka, Osaka 560-0043, Japan \\
$^2$Institute of Theoretical Physics Kanazawa University\\
Kakumamachi Kanazawa 920-1192, Japan 
}

\end{center}

\begin{abstract}
In this paper, we study three dimensional NL$\sigma$Ms within two kind of nonperturbative methods; WRG and large-$N$ expansion.
First, we investigate the renormalizability of some NL$\sigma$Ms using WRG equation.
We find that some models have a nontrivial UV fixed point and are renormalizable within nonperturbative method.
Second, we study the phase structure of $CP^{N-1}$ and $Q^{N-2}$ models using large-$N$ expansion.
These two models have two and three phases respectively.
At last, we construct the conformal field theories at the fixed point of the nonperturbative WRG $\beta$ function.

This is the review of recently works and is based on the talk of the conference by EI.
\end{abstract}

\section{Introduction}
The Wilsonian Renormalization Group (WRG) equation is a exact equation. 
That describes the variation of the general effective action when the cutoff scale is changed. However, the most general effective action depends on infinitely many coupling constants, and in practice, we have to introduce some kind of symmetry and truncation to solve the WRG equation. The simplest truncation is the local potential approximation, in which only the potential term without a derivative is retained. 
In this paper, we consider the next-to-leading order approximation, then such terms can be written as non-linear sigma models (NL$\sigma$Ms) action.
Since the higher derivative terms are irrelevant in the infrared (IR) region, we assume that we can safely drop such terms in an regions. Furthermore, we assume that the ${\cal N}=2$ supersymmetry is maintained in the renormalization procedure. In the ${\cal N}=2$ supersymmetric NL$\sigma$Ms, the field variables take values in complex curved spaces called K\"{a}hler manifolds, whose metrics are specified completely by the K\"{a}hler potentials. Because of ${\cal N}=2$ supersymmetry, non-derivative terms such as $\det{g}$, disappear through a remarkable cancellation between bosons and fermions, and our action is completely fixed by the K\"{a}hler potential. Thus, our truncation method in which higher derivative terms are ignored is the first nontrivial approximation for ${\cal N}=2$ supersymmetric NL$\sigma$M.

On the other hand, NL$\sigma$Ms are renormalizable in two dimensions. They are nonrenormalizable, however, in three dimensions within the perturbation theory. Therefore, we have to use nonperturbative methods to study the renormalizability of NL$\sigma$Ms. 
In this paper, we study three dimensional NL$\sigma$Ms within two kind of nonperturbative methods; WRG and large-$N$ expansion.

First, we investigate the renormalizability of some NL$\sigma$Ms
In the WRG approach, the renormalizability of NL$\sigma$Ms is equivalent to the existence of a nontrivial continuum limit, $\Lambda \rightarrow \infty$. When the ultraviolet (UV) cutoff $\Lambda$ tends to infinity, we have to fine tune the coupling constant to the critical value at the UV fixed point, so as to keep the observable quantities finite. Therefore, it is important to show the existence of the UV fixed point without using the perturbation theory. 
We show that if the target manifold has a positive scalar curvature, the theory has a nontrivial UV fixed point together with an IR fixed point at the origin. It is possible to take the continuum limit of this theory by using the UV fixed point, so that these NL$\sigma$Ms are renormalizable in three dimensions, at least within our truncation method.

Next, we study the phase structures of the C$P^{N-1}$ and $Q^{N-2}$ models using large-$N$ expansion.
From WRG study, we will find these two models has a nontrivial UV fixed point.
We will show that the $CP^{N-1}$ model has two phases: $SU(N)$ symmetric and broken phases, and the $Q^{N-2}$ model has three phases: Chern-Simons, Higgs and $SO(N)$ broken phases when $N$ is large region.

At last, we will construct the conformal field theories at the fixed point of the nonperturbative $\beta$ function in more general theory spaces. To simplify, we assume ${\bf SU}(N)$ symmetry for the theory and obtain a class of ${\bf SU}(N)$ invariant conformal field theories with one free parameter. If we choose a specific value of this free parameter, we obtain a conformal field theory defined at the UV fixed point of the C$P^N$ model and the symmetry is enhanced to ${\bf SU}(N+1)$ in this case.
The conformal field theory reduces to a free field theory if the free parameter is set to zero. Therefore, the free parameter describes a marginal deformation from the IR to the UV fixed point of the C$P^N$ model in the theory spaces.

\section{Wilsonian Renormalization Group (WRG) equation}\label{review}

In general, the WRG eq. describes the variation of the effective action $S$ when the cutoff energy scale $\Lambda$ is changed to $\Lambda (\delta t)=\Lambda e^{-\delta t}$ in $D$-dimensional field theory \cite{Wilson Kogut, Wegner and Houghton, Morris, Aoki}:
\beq  
\frac{d}{dt}S[\Omega; t]&=&\frac{1}{2\delta t} \int_{p'} tr \ln \left(\frac{\delta^2 S}{\delta \Omega^i \delta \Omega^j}\right)\nonumber\\
&&-\frac{1}{2 \delta t}\int_{p'} \int_{q'} \frac{\delta S}{\delta \Omega^i (p')} \left(\frac{\delta^2 S}{\delta \Omega^i (p')\delta \Omega^j (q')} \right)^{-1} \frac{\delta S}{\delta \Omega^j (q')} \nonumber\\
&&+ \left[D-\sum_{\Omega_i} \int_p \hat{\Omega}_i (p) \left(d_{\Omega_i}+\gamma_{\Omega_i}+\hat{p}^{\mu} \frac{\partial}{\partial \hat{p}^{\mu}} \right) \frac{\delta}{\delta \hat{\Omega}_i (p)} \right] \hat{S}.\nonumber\\ \label{WRG-1} 
\eeq
Here, $d_{\Omega}$ and $\gamma_{\Omega}$ denote the canonical and anomalous dimensions of the field $\Omega$, respectively.
The carat indicates dimensionless quantities.
The first and second terms in Eq.~(\ref{WRG-1}) correspond to the one-loop and dumbbell diagrams, respectively.
The remaining terms come from the rescaling of fields.
We always normalize the coefficient of the kinetic term to unity.

This WRG equation consists of an infinite set of differential equations for the various coupling constants within the most general action $S$.
In practice, we usually expand the action in powers of derivatives and retain the first few terms.
We often introduce some symmetry to further reduce the number of independent coupling constants.

We consider the simplest ${\cal N}=2$ supersymmetric theory, which is given by the K\"{a}hler potential term.
In this case, the action is written
\beq
S&=&\int d^2 \theta d^2 \bar{\theta} d^3 x K[\Phi, \Phi^\dag]\nonumber\\
&=&\int d^3 x \Bigg[g_{n \bar{m}}\left(\partial^{\mu} \varphi^n \partial_{\mu} \varphi^{* \bar{m}} +\frac{i}{2} \bar{\psi}^{\bar{m}} \sigma^{\mu}(D_{\mu} \psi)^n +\frac{i}{2} \psi^{n} \bar{\sigma}^{\mu}(D_{\mu} \bar{\psi})^{\bar{m}} +\bar{F}^{\bar{m}} F^{n}\right) \nonumber\\
&&-\frac{1}{2} K_{,nm \bar{l}} \bar{F}^{\bar{l}} \psi^n \psi^m -\frac{1}{2} K_{,n \bar{m} \bar{l}} F^{n} \bar{\psi}^{\bar{m}} \bar{\psi}^{\bar{l}}+\frac{1}{4} K_{,nm \bar{k} \bar{l}} (\bar{\psi}^{\bar{k}} \bar{\psi}^{\bar{l}})(\psi^n \psi^m)\Bigg],\label{action}
\eeq
where $\Phi^n$ represents chiral superfields, whose component fields are complex scalars $\varphi^n (x)$, Dirac fermions $\psi^n (x)$ and complex auxiliary fields $F^n (x)$.
In the usual perturbative method, this model is not renormalizable.
We examine it using the WRG equation, which constitutes one of nonperturbative methods.

When we substitute this action into Eq.(\ref{WRG-1}), the one-loop correction term cannot be written in covariant form \cite{HI}.
We use the K\"{a}hler normal coordinates (KNC) to obtain a covariant expression for the loop correction term. In order to integrate over the rapidly fluctuating fields, we have to expand the action around the slowly varying fields, which are treated as background fields. We transform the fluctuating fields to the KNC by holomorphic transformations. Then the action for the fluctuating fields is expressed in terms of covariant quantities under the general coordinate transformations of the background fields. After integration over the fluctuating fields, the resulting effective action for the background fields is covariant under the general coordinate transformations. 
The Jacobian for the path integral measure cancels between bosons and fermions because of the supersymmetry.
In nonlinear sigma models defined on a coset manifold $G/H$, a part of the global symmetry is realized nonlinearly. If the general coordinate transformation of the target manifold allows an infinitesimal transformation that leaves the metric invariant, the generator of the transformation is called a Killing vector and defines the symmetry of the corresponding nonlinear sigma model. In order to maintain the global symmetries that are realized nonlinearly, it is desirable to respect covariance under the general coordinate transformation.

Finally, we obtain the WRG equation for the scalar part as
\beq
&&\frac{d}{dt}\int d^3 x g_{i \bar{j}} (\partial_\mu \varphi)^i (\partial^\mu \varphi^*)^{\bar{j}}\nonumber\\
&&=\int d^3 x \Big[-\frac{1}{2 \pi^2} R_{i \bar{j}} \nonumber\\
&&-\gamma \Big(\varphi^k g_{i \bar{j},k}+ \varphi^{*\bar{k}}g_{i \bar{j},\bar{k}} +2g_{i \bar{j}} \Big) -\frac{1}{2} \Big( \varphi^k g_{i \bar{j},k}+ \varphi^{*\bar{k}}g_{i \bar{j},\bar{k}} \Big) \Big] (\partial_\mu \varphi)^i  (\partial^\mu \varphi^*)^{\bar{j}},\label{WRG3dim} \nonumber\\
\eeq
where it has been assumed that the scalar fields $\varphi^n (x)$ are made independent of $t$ through a suitable rescaling, which introduces the anomalous dimension $\gamma$.
From this WRG equation, the $\beta$ function of the K\"{a}hler metric is
\beq
\frac{d}{dt}g_{i \bar{j}}&=&-\frac{1}{2 \pi^2}R_{i \bar{j}} -\gamma \Big[\varphi^k g_{i \bar{j},k} +\varphi^{* \bar{k}}g_{i \bar{j},\bar{k}}+2g_{i \bar{j}} \Big] -\frac{1}{2} \Big[\varphi^k g_{i \bar{j},k} +\varphi^{* \bar{k}}g_{i \bar{j},\bar{k}} \Big]  \nonumber\\
&\equiv&-\beta (g_{i \bar{j}}).\label{beta}
\eeq

\section{Renormalizability of some models}\label{E-K}
Let us discuss the renormalizability of some theories whose target spaces are Einstein-K\"{a}hler manifolds.
The Einstein-K\"{a}hler manifolds satisfy the condition
\beq
R_{i \bar{j}}=\frac{h}{a^2}g_{i \bar{j}},\label{EKcond}
\eeq
where $a$ is the radius of the manifold, which is related to the coupling constant $\lambda$ by
\beq
\lambda =\frac{1}{a}.
\eeq
There is a special class of Einstein-K\"{a}hler manifolds called the Hermitian symmetric space that consists of a symmetric coset space ($G/H$), namely, if the coset space is invariant under a parity operation. If the manifold is the Hermitian symmetric space, the positive constant $h$ in Eq.~(\ref{EKcond}) is the eigenvalue of the quadratic Casimir operator in  the adjoint representation of the global symmetry $G$, as shown in Table $1$.

	
	\begin{table}[h]
	
	\begin{center}
	\begin{tabular}{|c|c|c|}
	\hline
	$G/H$                              & Dimensions (complex)  &$h$\\
	\hline \hline
	${\bf SU}(N)/[{\bf SU}(N-1) \otimes {\bf U}(1)]=$C$P^{N-1}$   & $N-1$        & $N$\\
	${\bf SU}(N)/[{\bf SU}(N-M) \otimes {\bf U}(M)]$   & $M(N-M)$                & $N$\\
	${\bf SO}(N)/[{\bf SO}(N-2) \otimes {\bf U}(1)]=Q^{N-2}$   & $N-2$           & $N-2$\\
	${\bf Sp}(N)/{\bf U}(N)$                     & $\frac{1}{2}N(N+1)$   & $N+1$\\
	${\bf SO}(2N)/{\bf U}(N)$                    & $\frac{1}{2}N(N+1)$   & $N-1$\\
	$E_{6}/[{\bf SO}(10) \otimes {\bf U}(1)]$    &$16$                     &$12$\\
	$E_{7}/[E_6 \otimes {\bf U}(1)]$        &$27$                     &$18$\\
	\hline
	\end{tabular}
	\caption{The values of $h$ for Hermitian symmetric spaces}\label{table-HHS}
	\end{center}
	\end{table}
	

To obtain the anomalous dimension of the scalar fields and the $\beta$ function of the coupling constant $\lambda$, we substitute the Einstein-K\"{a}hler condition (\ref{EKcond}) for the WRG $\beta$ function (\ref{beta}).
Because only $\lambda$ depends on $t$, we obtain the anomalous dimension of scalar fields (or chiral superfields) as
\beq
\gamma=- \frac{h \lambda^2}{4\pi^2}.\label{gamma},
\eeq
and the $\beta$ function of $\lambda$:
\beq
\beta(\lambda)&\equiv&-\frac{d \lambda}{dt}=-\frac{h}{4\pi^2}\lambda^3+\frac{1}{2} \lambda .\label{beta-lambda}
\eeq

For positive $h$, we have an IR fixed point at 
\beq
\lambda=0,
\eeq
and we also have a UV fixed point at 
\beq
\lambda^2=\frac{2 \pi^2}{h}\equiv \lambda_c^2.
\eeq
{\it Therefore, if the constant $h$ is positive, it is possible to take the continuum limit by choosing the cutoff dependence of the bare coupling constant as
\beq
\lambda(\Lambda) \stackrel {\Lambda \rightarrow \infty}{\longrightarrow} \lambda_c-\frac{M}{\Lambda},\label{continuum}
\eeq
where $M$ is a finite mass scale.}
With this fine tuning, ${\cal N}=2$ supersymmetric nonlinear $\sigma$ models are renormalizable, at least in our approximation, if the target spaces are Einstein-K\"{a}hler manifolds with positive curvature.

Furthermore when the constant $h$ is positive, the target manifold is a compact Einstein-K\"{a}hler manifold.
In this case, the anomalous dimensions at the fixed points are given by
\beq
\gamma_{IR}&=&0 \mbox{ at the IR fixed point (Gaussian fixed point),}\\
\gamma_{UV}&=&-\frac{1}{2} \mbox{at the UV fixed point.}
\eeq
At the UV fixed point, the scaling dimension of the scalar fields ($x_{\varphi}$) is equal to the canonical plus anomalous dimensions:
\beq
x_{\varphi}&\equiv& d_{\varphi} + \gamma_{\varphi}=0.
\eeq
Thus the scalar fields and the chiral superfields are dimensionless in the UV conformal theory, as in the case of two dimensional field theories.
Above the fixed point, the scalar fields have non-vanishing mass, and the symmetry is restored \cite{HKNT,HIT}.

On the other hand, for negative $h$, there is only IR fixed point at $\lambda=0$.
When the cutoff $\Lambda$ becomes large, the coupling constant goes to infity at finite $\Lambda$. Then if the constant $h$ is negative, the theories are nonrenormalizable within nonperturbative method.
\section{Phase structure of ${\bf C}P^{N-1}$ and $Q^{N-2}$ models}\label{phase}
In previous section, we find there might be the phase transition at nontrivial UV fixed point, if the constant $h$ is positive.
In this section, we study the phase structure for such two examples, C$P^{N-1}$ and $Q^{N-2}$ models which are the first and third line in Table \ref{table-HHS} respectively.
To study that, we use the large-$N$ expansion which is the other nonperturbative method {\cite{HIT, Inami}}.
\subsection{The ${\bf C}P^{N-1}$ model} \label{sec-CPN}
\subsubsection{The auxiliary field formulation of ${\bf C}P^{N-1}$ model}
Let us introduce chiral superfields $\Phi^i(x,\theta)\ (i=1,2, \cdots, N)$ belonging to a fundamental representation of $G=SU(N)$, the isometry group of ${\bf C}P^{N-1}$. 
We also introduce $U(1)$ gauge symmetry 
\begin{equation}
 \Phi(x,\theta) \longrightarrow \Phi'(x,\theta)
  = e^{i \Lambda(x,\theta)} \Phi(x,\theta) \label{eqn:projective}
\end{equation}
to require that $\Phi(x,\theta)$ and $\Phi'(x,\theta)$ are 
physically indistinguishable. 
With a complex chiral superfield, 
$e^{i \Lambda(x,\theta)}$ is an arbitrary complex number. 
$U(1)$ gauge symmetry is thus complexified to $U(1)^{\bf C}$. 
The identification 
$\Phi \sim \Phi'$ defines the complex projective space ${\bf C}P^{N-1}$.
In order to impose local $U(1)$ gauge symmetry, 
we have to introduce a $U(1)$ gauge field 
$V(x, \theta, \bar{\theta})$ with 
the transformation property 
$e^{-V} \longrightarrow e^{-V} e^{-i\Lambda + i\Lambda^*}$.
$V(\theta,\bar{\theta},x)$ is a real scalar superfield\cite{WB}, and defined by dimensional reduction from $4$-dimensional ${\cal N}=1$ to $3$-dimensions. 
Then the Lagrangian with a local $U(1)$ gauge symmetry is given by
\beq
{\cal L}=\int d^4 \theta (\Phi^i \Phi^{\dag i} e^{-V} +c V), \label{CPN-kahler2}
\eeq
where the last term $V$ is called the Fayet-Illiopoulos D-term.
In this model, the gauge field $V(x,\theta,\bar{\theta})$ 
is an auxiliary superfield without kinetic term. 
The K\"ahler potential $K(\Phi, \Phi^\dag)$ 
is obtained by eliminating $V$ 
using the equation of motion for $V$
\begin{equation}
  {\cal L} = \int d^2\theta d^2\bar{\theta} K(\Phi,\Phi^\dag) 
  = c\int d^2\theta d^2\bar{\theta} 
     \log{({\vec{\Phi}}^\dag \cdot\vec{\Phi})}.
  \label{eqn:kahlerpot}
\end{equation}
This K\"ahler potential reduces to the standard Fubini-Study metric of 
${\bf C}P^{N-1}$ 
\begin{equation}
  K(\Phi,\Phi^\dag)=c\log{(1+\sum_{i=1}^{N-1}\Phi^{\dag i} \Phi^i)}
   \label{eqn:fsmetric}
\end{equation}
by a choice of gauge fixing
\begin{equation}
  \Phi^N(x,\theta)=1.\label{eqn:cpgauge}
\end{equation}
The global symmetry $G=SU(N)$, the isometry of the target space
${\bf C}P^{N-1}$, is linearly realized on our $\Phi^i$ fields and our 
Lagrangian (\ref{CPN-kahler2}) with auxiliary field $V$ is manifestly 
invariant under $G$. The gauge fixing condition (\ref{eqn:cpgauge}) is
not invariant under $G=SU(N)$ and we have to perform an appropriate 
gauge transformation simultaneously to compensate the change of
$\Phi^N$ caused by the $SU(N)$ transformation. Therefore the global 
symmetry $G=SU(N)$ is nonlinearly realized in the gauge fixed theory. 
In this sense, our Lagrangian (\ref{CPN-kahler2}) use the linear 
realization of $G$ in contrast to the nonlinear Lagrangian (\ref{eqn:fsmetric}) in terms 
of the K\"ahler potential which use the nonlinear realization of $G$.

Instead of fixing the gauge to eliminate one component of chiral superfield, we can eliminate the chiral and and anti-chiral components in the gauge superfield $V(x,\theta,\bar{\theta})$. In our discussion below, we employ this Wess-Zumino gauge. Then the gauge superfield is written by component fields:
\beq
V=\bar{\theta} \gamma^\mu \theta v_{\mu} +\bar{\theta} \theta M +\frac{1}{2}\theta^2\bar{\theta}\lambda +\frac{1}{2} \bar{\theta}^2\bar{\lambda}\theta  + \frac{1}{4}\theta^2\bar{\theta}^2D, \nonumber
\eeq
where $v_{\mu}$ is a gauge field of three dimensional theory and the scalar field $M$ corresponds to the fourth component of $4$-dimensional vector field. Note that the real part of the scalar component of the gauge transformation $\Lambda(x,\theta)$ is not fixed yet with this gauge choice. We have to fix this residual gauge freedom to calculate the $1/N$ correction where this auxiliary gauge field begins to propagate. 
In order to make the Lagrangian of order $N$ in the large $N$ limit, we take the coefficient Fayet-Illiopoulous D-term $c=N/g^2$ and keep $g^2$ fixed when we take the limit of $N \rightarrow \infty$.
With this gauge choice, the global $SU(N)$ symmetry coming from the isometry of the target manifold is linearly realized, while the supersymmetry is realized non-linearly.
Furthermore, the action is invariant under $U(1)$ gauged symmetry generated by the real part of the gauge function $\Lambda(x,\theta)$, with the assignment of the $U(1)$ charge
\beq
[\Phi]=1 , \hspace{0.5cm} [\Phi^\dag]=-1.\label{U(1)sym}
\eeq

Integrating out the Grassmann coordinates $\theta$ and $\bar{\theta}$, the action (\ref{CPN-kahler2}) is written by using the component fields:
\beq
{\cal L}&=&\partial_\mu \varphi^{*i } \partial^\mu \varphi^i +i \bar{\psi}^i \partial\llap / \psi^i +F^i F^{i*} -[i (\varphi^{*i} \partial_\mu \varphi^i -\varphi^i \partial_\mu \varphi^{i*}) +\bar{\psi}^i \gamma_\mu \psi^i ]v^\mu +v^\mu v_\mu \varphi^{*i } \varphi^i \nonumber\\
&& - M^2 \varphi^{*i } \varphi^i -M \bar{\psi}^i \psi^i -D \varphi^{*i} \varphi^i +\frac{N}{g^2} D +(\varphi^i \bar{\psi}^i\lambda +\varphi^{i*} \bar{\lambda} \psi^i)\label{lagrangian-CPN}
\eeq
Since the gauge superfield $V$ or its component fields do not have kinetic terms, they are auxiliary fields and do not propagate in the tree approximation.
If we eliminate all auxiliary fields using their equations of motion, we obtain the constraints
\beq
&&\varphi^i \varphi^{* i}=\frac{N}{g^2}, \quad 
M=-\frac{g^2}{2N} \bar{\psi}^{i}\psi^i, \nonumber\\
&&v^\mu=\frac{g^2}{2N}\left[i (\varphi^{*i} \partial_\mu \varphi^i -\varphi^i \partial_\mu \varphi^{i*})+ \bar{\psi}^{i}\gamma^\mu \psi^i\right], \label{eom-CPN}\\
&&\varphi^i \bar{\psi}^i =\varphi^{*i} \psi^i=0. \nonumber
\eeq
The first equation means fields $\varphi^i$ are constrained on the $(2N-1)$ dimensional sphere $S^{2N-1}$.
Furthermore, the gauge transformation of gauge field $v_\mu$ eliminates a common phase of $\varphi^i$.
Thus, the target manifold reduces to the complex projective space ${\bf C}P^{N-1}$. The last equation of (\ref{eom-CPN}) implies that fermion resides on the tangent space of ${\bf C}P^{N-1}$.

\subsubsection{Phase structure of ${\bf C}P^{N-1}$ model}
To investigate the phase structure, let us calculate the effective potential in the leading order of the $1/N$ expansion.
The partition function of this model can be written as
\beq
Z=\int D\Phi^i D \Phi^{\dag \bar{i}} DV e^{i \int d^3 x {\cal L}},
\eeq 
where the Lagrangian ${\cal L}$ is given in (\ref{CPN-kahler2}) and the measure $DV$ includes the gauge fixing term of the remaining gauge freedom in the Wess-Zumino gauge. In order to calculate the effective potential, the vacuum energy when $\langle \varphi(x)\rangle=\tilde{\varphi}$ is kept fixed,  
we divide the dynamical field into the vacuum expectation value and the fluctuation, $\varphi^i=\tilde{\varphi}^i +\varphi^{'i}$.
The fluctuation field has to follow a constraint \cite{HKNT,KH}
\beq
\int d^3 x \varphi^{'i}(x)=0,
\eeq
which forbids the appearance of tadpole or one particle reducible graphs in the effective potential.
Integrating over the fluctuation field $\varphi^{'i}$, we obtain the effective action:
\beq
Z&=&\int DV e^{i S_{eff}},\label{partition_fn}\\
S_{eff}&=&-\frac{N}{i}{\rm Tr} \ln (\nabla_{B} +\bar{\lambda} \nabla^{-1}_F \lambda) +\frac{N}{i}{\rm Tr} \ln \nabla_{F}\label{eff-action} \\
&&+\int d^3x\left(\frac{N}{g^2} D - \tilde{\varphi}^{*i }(M^2+D) \tilde{\varphi}^i+ ( \mbox{$\lambda$-dependent terms})\right). \nonumber
\eeq
Here we use the following notation
\beq
\nabla_{B}=D^\mu D_\mu +(M^2 +D),\hspace{0.5cm} \nabla_{F}=i \gamma^{\mu}D_{\mu}-M,\hspace{0.5cm} 
D_\mu = \partial_\mu +i v_\mu. \nonumber
\eeq
Assuming the Lorentz invariance of the vacuum, we take the vacuum expectation value of each auxiliary field as follow:
\beq
\langle M(x) \rangle =M_0 , \hspace{0.5cm} \langle D(x) \rangle=D_0, \hspace{0.5cm} \langle {\mbox{others}} \rangle=0. \nonumber
\eeq
We evaluate the momentum integration by introducing the ultraviolet cutoff $\Lambda$, then we obtain the effective potential in the leading order of the $1/N$ expansion ($k\llap /=\gamma^{\mu}k_{\mu}$).
\beq
\frac{V_{\rm eff}}{N}
&=&-\frac{1}{6 \pi} |M_0^2+D_0|^{\frac{3}{2}} +\frac{1}{6\pi}|M_0|^3 +\frac{1}{N}(M_0^2 +D_0)|\tilde{\varphi}^i|^2 +\frac{m}{4 \pi} D_0, \label{CPN-potential} 
\eeq
where we defined a renormalized coupling constant $g_R$ to absorb the linear divergence that appears in the coefficient of $D_0$
\beq
\frac{\mu}{g_R^2}&=&\frac{1}{g^2}-\frac{1}{2\pi^2}\Lambda +\frac{\mu}{4 \pi}.\label{renormalized_coupling}
\eeq
with an arbitary finite mass scale $\mu$ called the renormalization point. We have also defined a renormalization group invariant mass $m$ given by 
\beq
m &\equiv& \mu (1-\frac{4 \pi}{g_R^2}). \label{mass}
\eeq

To evaluate the effective potential in the presence of $\tilde{\varphi}, M_0, D_0$, we have to perform the path integration over the nonzero mode of auxiliary fields in (\ref{partition_fn}). When $N$ is very large, these path integration can be done by the saddle point method since the $S_{eff}$ is of order $N$. In the leading order of $1/N$ expansion, the effective potential is given by the value of $S_{eff}$ at the saddle point. Assuming the saddle point is located at the translationally invariant configuration, we can neglect the non-zero mode of the auxiliary fields in the large $N$ limit. The saddle point condition is, therefore, equivalent to the the stationary condition of the effective potential (\ref{CPN-potential}): 
These two conditions fixes the value of $M_0$ at the saddle point
\begin{equation}
|M_0|=0 \quad \mbox{or}\quad m. \label{m0_saddle}
\end{equation}
We will discuss these two cases separately.
\begin{description}
\item[$|M_0|=m$ case:]
This case is possible only when $m\ge 0$. By solving the stationary condition of the effective potential, we obtain the vacuum energy when $\tilde{\varphi}$ is kept fixed
\begin{equation}
V_{\rm eff}(\tilde{\varphi})=\frac{N}{12\pi}\left(\left|\frac{4\pi}{N}|\tilde{\varphi}^i|^2+m\right|^3-m^3\right).\label{v1_eff}
\end{equation}
Because we are assuming $m\ge 0$, the minimum of this vacuum energy is located at $\tilde{\varphi}^i=0$ and $D_0=0$. Since $\tilde{\varphi}$ and $D_0$ are the order parameter of $SU(N)$ and supersymmetry respectively, neither $SU(N)$ nor supersymmetry is broken in this case.

\item[$M_0=0$ case:]
By solving the stationary condition of the effective potential, we obtain the vacuum energy
\begin{equation}
V_{\rm eff}(\tilde{\varphi})=\frac{N}{12\pi}\left|\frac{4\pi}{N}|\tilde{\varphi}^i|^2+m\right|^3.\label{v2_eff}
\end{equation}
When $m\ge 0$, the minimum located at $\tilde{\varphi}=0$ has a higher vacuum energy than the previous vacuum, and does not correspond to the ground state.
On the other hand, when $m<0$, the minimum is at 
\begin{equation}
|\tilde{\varphi}^i|^2=\frac{N}{4\pi}|m|,\label{vev_broken}
\end{equation}
and $D_0=0$, namely supersymmetry is unbroken while $SU(N)$ symmetry is spontaneously broken in this case.
\end{description}

Thus we found two vacuum, both of them are supersymmetric. In the following, we will discuss these two phases in detail\cite{HIT}. 

\begin{enumerate}
\item $SU(N)$ symmetric phase  ($|\tilde{\varphi}^i|^2=0,  M_0 =m$)\\
We find that $N$ scalar and spinor fields have the same mass $m$ by the vacuum expectation value of $M_0$, and global SU(N) and local $U(1)$ gauge symmetry (\ref{U(1)sym}) are unbroken in this phase.
Integrating out the dynamical field, we obtain the effective action (\ref{eff-action}) describing the dynamics of auxiliary fields in this phase.

From the effective action, we can find the Chern-Simons interaction term is induced in this phase.
The gauge field $v_\mu$ acquires mass through this term. 
Calculating the two point functions of all auxiliary fields, we find the mass of the fields $v_{\mu}, M, \lambda$ is twice of that of the dynamical fields. These auxiliary field $v_{\mu}, M$ and $\lambda$ represent the bound states of fermion--anti-fermion and fermion--boson as is shown in (\ref{eom-CPN}). 

\item $SU(N)$ broken phase  ($|\tilde{\varphi}^i|^2=\frac{N}{4 \pi}|m|,  M_0 =0$)\\
Using $SU(N)$ symmetry, we choose the vacuum expectation value of $\tilde{\varphi}$ in the $N$-th direction: $\tilde{\varphi}^N=\sqrt{\frac{N}{4 \pi}|m|}$. 
In this phase, global $SU(N)$ symmetry is broken down to $SU(N-1)$ and there are $N-1$ massless Nambu-Goldstone bosons $\varphi^i\ (i=1,\cdots,N-1)$, and their superpartners $\psi^i\ (i=1,\cdots,N-1)$.
Furthermore, gauged $U(1)$ symmetry (\ref{U(1)sym}) is also broken because the dynamical fields $\varphi^N$, carrying $U(1)$ charge, have a nonvanishing vacuum expectation value.
We find all auxiliary fields are not physical particles in the broken phase.
\end{enumerate}

From the eq.(\ref{mass}), the region where the coupling constant is smaller than the critical point $g_c$ corresponds to broken phase.

\begin{figure}[h]
\begin{center}

\unitlength=1mm
\begin{picture}(50,40)

\includegraphics[width=5cm]{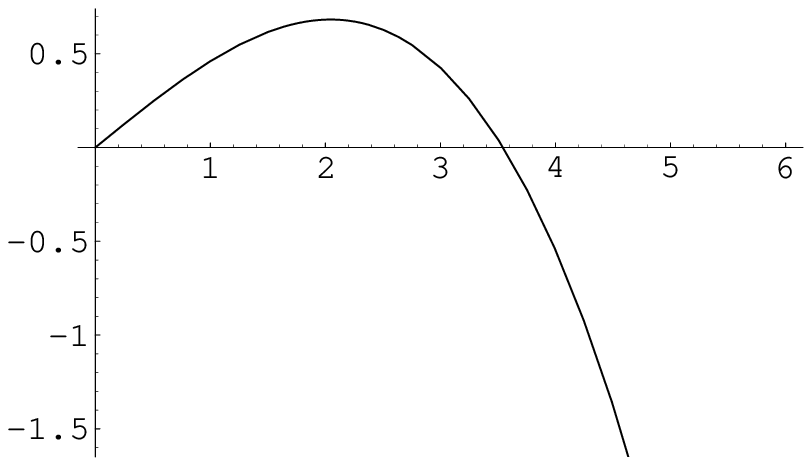}
\put(-51,30){\bf $\beta (g)$}
\put(-20,17) {\bf$g_c$}
\put(0,17){\bf $g$}
\put(-15,24){\bf The symmetric and}
\put(5,20) {\bf massive phase}
\put(-42,0){\bf The broken and} 
\put(-42,-4){\bf massless phase}

\end{picture}
\caption{The $\beta$ function of the coupling constant $g$ for the $CP^{N-1}$ model.}\label{3dim-beta}

\end{center}

\end{figure}
\subsection{The $Q^{N-2}$ model}\label{sec-QN}
\subsubsection{The theory and symmetry}
Let's investigate $Q^{N-2}$ model using $1/N$ expansion.
This model is obtained from $CP^{N-1}$ model with the $O(N)$ condition $\Phi^i \Phi^i =0$.
Then the Lagrangian has the two kinds of auxiliary fields as follow:
\beq
{\cal L}=\int d^4 \theta (\Phi^i \Phi^{\dag \bar{i}} e^V -c V)+\frac{1}{2} \Big( \int d^2 \theta \Phi_0 \Phi^i \Phi^i +\int d^2 \bar{\theta} \Phi^\dag_0 \Phi^{\dag i } \Phi^{\dag i}   \Big),
\eeq
where $V(\theta,\bar{\theta},x)$ is $U(1)$ gauge superfield and $\Phi_0 ,\Phi^{\dag}_0$ are chiral- and antichiral- superfields respectively.
Components of the new auxiliary field $\Phi_0$ is defined by
\beq
\Phi_0(y)&=&A_0 (y)+\sqrt{2} \theta \psi^i_0 (y) +\theta \theta F_0 (y).\nonumber
\eeq
This auxiliary superfield is an $O(N)$ singlet and has gauged nonzero $U(1)$ charge; $[\Phi_0]=-2$ and $[\Phi_0^\dag]=2$.
The dynamical fields $\Phi^i$ is a vector representation of global $O(N)$ symmetry and has the gauged $U(1)$ charge (\ref{U(1)sym}).
Then the $Q^{N-2}$ model has global $SO(N)$ symmetry coming from the isometry of the target manifold and the gauged $U(1)$ symmetry.

Integrated out the Grassmann coordinates $\theta$ and $\bar{\theta}$, we obtain the $Q^{N-2}$ model Lagrangian in component fields:
\beq
{\cal L}&=&\partial_\mu \varphi^{*i } \partial^\mu \varphi^i +i \bar{\psi} \partial_\mu \gamma^\mu \psi^i +F^i F^{*i} -[i (\varphi^{*i } \partial_\mu \varphi^i -\varphi^i \partial_\mu \varphi^{*i}) -\bar{\psi}^i \gamma_\mu \psi^i ]v^\mu +v^\mu v_\mu \varphi^{*i } \varphi^i \nonumber\\
&& - M^2 \varphi^{*i } \varphi^i -M \bar{\psi}^i \psi^i -D \varphi^{*i} \varphi^i +\frac{N}{g^2} D +(\varphi^i \bar{\psi}^i \lambda ^c +\varphi^{*i} \lambda^c \psi^i)\nonumber\\
&&+\frac{1}{2}(F_0 \varphi^i \varphi^i +F_0^{*} \varphi^{* i} \varphi^{* i})-(\bar{\psi}_0^c \psi^i \varphi^i +\bar{\psi}^i \psi_0^c \varphi^{* i}) \nonumber\\
&&+A_0 (\varphi^i F^i -\frac{1}{2} \bar{\psi}^{c i} \psi^i)
+A_0^*(\varphi^{*i} F^{*i} -\frac{1}{2} \bar{\psi}^{ i} \psi^{c i}).\label{QN-Lagrangian}
\eeq
If we eliminate auxiliary fields using their equation of motion, we obtain the constraint eqs.(\ref{eom-CPN}) and additional following equations.
\beq
\varphi^i \varphi^i=\varphi^{*i} \varphi^{*i}=0, \hspace{0.5cm} 
\varphi^i \psi^i=\varphi^{*i} \bar{\psi}^{i}=0, \hspace{0.5cm}
A_0=-\frac{g^2}{2N} \bar{\psi}^{c i}\psi^{i}, \hspace{0.5cm}
A_0^{*}=-\frac{g^2}{2N} \bar{\psi}^{ i}\psi^{ci}.\label{eom-QN}
\eeq
The equations (\ref{eom-QN}) is similar to ${\cal N}=1$ supersymmetric $O(N)$ model with zero radius.

\subsubsection{Phase structure}
Similarly to $CP^{N-1}$ case, we integrate out the fluctuation field and put the vacuum expectation values as follows:
\beq
\langle M(x) \rangle &=&M_0, \hspace{0.5cm} \langle D(x) \rangle=D_0,  \nonumber\\
\langle F_0(x) \rangle &=&F_0 , \hspace{0.5cm} \langle A_0(x) \rangle=A_0, \hspace{0.5cm} \langle \mbox{the others} \rangle=0. \nonumber
\eeq
Then we obtain the effective potential.
\beq
\frac{V}{N}
&=&-\frac{1}{6 \pi} \frac{1}{2} \Big[|M_0^2+D_0+|A_0|^2 +|F_0||^{\frac{3}{2}}+|M_0^2+D_0+|A_0|^2 -|F_0||^{\frac{3}{2}} \Big]\nonumber\\
&& +\frac{1}{6\pi}\frac{1}{2}\Big[ |M_0+|A_0||^3 +|M_0-|A_0||^3 \Big] +\frac{m}{4 \pi}D_0 \nonumber\\
&&+\frac{1}{N} \Big(M_0^2 \varphi^{*i} \varphi^i +A_0^* A_0 \varphi^{*i}\varphi^i +D_0 \varphi^{*i}\varphi^i -\frac{1}{2} (F_0 \varphi^i \varphi^i +F_0^* \varphi^{*i}\varphi^{*i}) \Big).\nonumber
\eeq
We use the same renormalization of coupling constant and the same invariant mass as $CP^{N-1}$ case.

To obtain the vacuum expectation values at the stationary points of this effective potential, we will discuss two cases , $m>0$ and $m<0$, separately.

\begin{description}
\item[$m>0$ case:]
By solving the stationary conditions of the effective potential, we find the vacuum expectation values of the $M_0$ and $A_0$ as follow:
\beq
M_0=m, A_0=0 \hspace{0.5cm} \mbox{or} \hspace{0.5cm}M_0=0, |A_0|=m.
\eeq
Substituting these values back to the effective potential, we obtain the vacuum energy
\beq
V_{eff}=\frac{1}{24 \pi} \Bigg(|\frac{4 \pi}{N} \varphi_1^2|^3 +|\frac{4 \pi}{N}\varphi_2^2 +m|^3  \Bigg) -\frac{1}{12 \pi} m^3,
\eeq
where we put $\varphi^i=\frac{1}{\sqrt{2}}(\varphi_1 +i\varphi_2)$.
The minimum of this effective potential is located at $\varphi_1=\varphi_2=0$ and $D_0=0$.
We call the vacuum with $M_0=m, A_0=0$ the {\bf Chern-Simons phase}, and the vacuum with $M_0=0, |A_0|=m$ the {\bf Higgs phase} respectively.
In both phases, neither $SO(N)$ nor supersymmetry is broken because $\varphi_1$, $\varphi_2$ and $D_0$ are zero which are the order parameter of $SO(N)$ and supersymmetry respectively.

\item[$m<0$ case:]
In this case, the stationary conditions give following relations:
\beq
M_0=&A_0&=0 \nonumber\\
|D_1|^{\frac{1}{2}} = \frac{4 \pi}{N} \varphi_1^2 +m, &\hspace{0.5cm}& |D_2|^{\frac{1}{2}}= \frac{4 \pi}{N} \varphi_2^2 +m,\label{QN-vev}
\eeq
where we put $D-|F_0|=D_1, D+|F_0|=D_2$.
Substituting these relations into the effective potential, we obtain the vacuum energy 
\beq
V_{eff}=\frac{1}{24 \pi} \Bigg( |\frac{4 \pi}{N} \varphi_1^2 +m |^3 + |\frac{4 \pi}{N} \varphi_2^2 +m| \Bigg).
\eeq
Since $m<0$ in this case, the minimum of this energy is located at 
\beq
\varphi_1^2 = \frac{N}{4 \pi} |m|, \hspace{0.5cm} \varphi_2^2=\frac{N}{4 \pi}|m|.
\eeq
From eq.(\ref{QN-vev}), we find $D_0=0$ and the supersymmetry is preserved. 
On the other hand, $SO(N)$ symmetry is broken because of the nonzero vacuum expectation values of the dynamical fields.

\end{description}

Thus we found three kinds of phases, all phases are supersymmetric.
In the following, we will discuss these three phases in detail\cite{HIT}.
\begin{enumerate}
\item Chern-Simons phase ($M_0=m$, $|A_0|=0$, $D=F_0=\varphi_1=\varphi_2=0$.)\\
This phase is very similar to the symmetric phase of $CP^{N-1}$.
We replace $M$ by $m+M'$ in the Lagrangian (\ref{QN-Lagrangian}), and we find that $N$ scalar and spinor fields have same mass $m$ because of the vacuum expectation value of $M_0$.
Global $SO(N)$ and gauged $U(1)$ (\ref{U(1)sym}) are unbroken in this phase.

In this phase, the Chern-Simons interaction term is induced.
The gauge field $v_\mu$ acquires mass through this term, and the local $U(1)$ gauge symmetry is broken.
We find the mass of the fields $v_\mu$, $M$, $\lambda$, $A_0$, $\psi_0$ is twice of that the dynamical fields.
The auxiliary field with the vacuum expectation value $M$ corresponds to the bound state of the fermion and anti-fermion.

\item Higgs phase ($M_0=0$, $|A_0|=m$, $D=F_0=\varphi_1=\varphi_2=0$.)\\
In this phase, global $SO(N)$ symmetry is protected and $N$ dynamical fields have masses due to the vacuum expectation value of $A_0$.
	However gauged $U(1)$ symmetry (\ref{U(1)sym}) is broken because the superfields $\Phi_0$, which has nonzero $U(1)$ charge, has nonvanishing vacuum expectation value.

From the propagators of the axiliary fields, we find the fields $M$ and $A_R$ has twice masses of dynamical fields.
From eq. of motion for $A_0$, we find the fields $A_0^*$ corresponds to the pair of fermion and fermion:
\beq
A_0^*=\frac{g^2}{4N} \bar{\psi}^{ci} \psi^i.
\eeq 
Furthermore, we found that the gauge bosons acquire masses through the Higgs mechanism and the imaginary part of $A_0$ is removed from the theory.

\item broken phase ($M_0=|A_0|=D=F_0=0$, $\frac{4 \pi}{N}\varphi_1^2+m=0$, $\frac{4 \pi}{N}\varphi_2^2 +m=0$($m<0$))\\
Using $SO(N)$ symmetry and the constraint $\varphi_1 \varphi_2=0$ which is obtained by the stationary condition, we choose the vacuum expectation values as follow:
\beq
      \varphi_{N-1}=i \sqrt{\frac{N}{4 \pi}|m|},
      \varphi_N  =\sqrt{\frac{N}{4 \pi}|m|}, 
      \mbox{the others}=0
\eeq
Thus global $SO(N)$ symmetry breaks to $SO(N-2)$, and there are $N-2$ massless Nambu-Goldstone bosons and their superpartners.
We find all auxiliary fields are not physical in this phase.

\end{enumerate}

\begin{figure}[h]
\begin{center}

\unitlength=1mm
\begin{picture}(50,40)

\includegraphics[width=5cm]{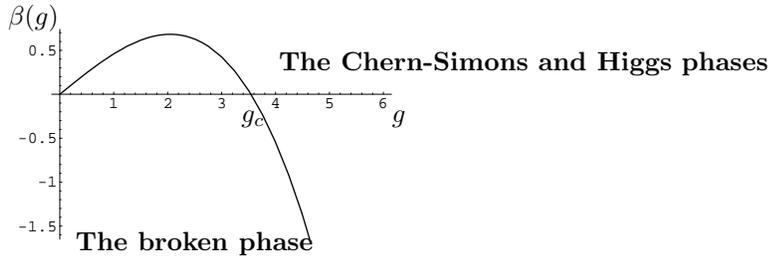}
\put(-51,30){\bf $\beta (g)$}
\put(-20,17) {\bf$g_c$}
\put(0,17){\bf $g$}
\put(-15,24){\bf The Chern-Simons and Higgs phases}
\put(-42,0){\bf The broken phase}

\end{picture}
\caption{The $\beta$ function of the coupling constant $g$ for $Q^{N-2}$ model.}\label{QN-beta-fig}

\end{center}

\end{figure}

Similarly to the $CP^{N-1}$ model, the region where the coupling constant is smaller than the critical point $g_c$ corresponds to broken phase, while the region where the coupling constant is larger than $g_c$ corresponds to Chern-Simons and Higgs phase.

\section{The SU(N) symmetric solution of the WRG equation}\label{solution}
In this section, we investigate the conformal field theories defined as the fixed points of the WRG $\beta$ function
\beq
\beta&=&\frac{1}{2 \pi^2}R_{i \bar{j}} +\gamma \Big[\varphi^k g_{i \bar{j},k} +\varphi^{* \bar{k}}g_{i \bar{j},\bar{k}}+2g_{i \bar{j}} \Big] +\frac{1}{2} \Big[\varphi^k g_{i \bar{j},k} +\varphi^{* \bar{k}}g_{i \bar{j},\bar{k}} \Big]\nonumber\\
&=&0.\label{beta2} 
\eeq
To simplify, we assume the ${\bf S}{\bf U}(N)$ symmetric K\"{a}hler potential
\beq
K[\Phi,\Phi^\dag]&=&\sum_{n=1}^{\infty} g_n (\vec{\Phi} \cdot \vec{\Phi}^\dag)^n \equiv f(x),\label{potential}
\eeq
where the chiral superfields $\vec{\Phi}$ have $N$ components, $g_n$ plays the role of the coupling constant, and $x \equiv \vec{\Phi} \cdot \vec{\Phi}^\dag$.
Using the function $f(x)$, we derive the K\"{a}hler metric and Ricci tensor as follows:
\beq
g_{i \bar{j}}&\equiv&\partial_i \partial_{\bar{j}} K[\Phi,\Phi^\dag]=f' \delta_{i \bar{j}}+f'' \varphi_i^* \varphi_{\bar{j}},\label{metric}\\
R_{i \bar{j}}&\equiv&-\partial_i \partial_{\bar{j}} \tr \ln g_{i \bar{j}}\nonumber\\
&=&-\Big[(N-1)\frac{f''}{f'} +\frac{2f''+f''' x}{f'+f'' x}  \Big]\delta_{i \bar{j}} \nonumber\\
&&-\Big[(N-1)\bigg(\frac{f^{(3)}}{f''}-\frac{(f'')^2}{(f')^2} \bigg)+\frac{3f^{(3)}+f^{(4)} x}{f'+f'' x} -\frac{(2f''+f'''x)^2}{(f'+f''x)^2} \Big]\varphi^*_{i}\varphi_{\bar{j}},\nonumber\\
\eeq
where 
\beq
f'=\frac{df}{dx}.
\eeq
To normalize the kinetic term, we set 
\beq
f'|_{x \approx 0}=1
\Rightarrow g_1=1. \label{normalization}
\eeq

To obtain a conformal field theory, we substitute the above metric and Ricci tensor into the $\beta$ function (\ref{beta2}) and solve the differential equation
\beq
\frac{\partial}{\partial t}f'=\frac{1}{2\pi^2}\Big[(N-1)\frac{f''}{f'} +\frac{2f''+f''' x}{f'+f'' x}  \Big]- 2\gamma(f'+f''x)\label{f'}-f''x =0\label{beta=0}.
\eeq
The function $f(x)$ is a polynomial of infinite degree, and it is difficult to solve it analytically. For this reason, we truncate the function $f(x)$ at order $O(x^4)$.
From the normalization (\ref{normalization}), the function $f(x)$ is 
\beq
f(x)=x+g_2 x^2 +g_3 x^3 +g_4 x^4.\label{order4}
\eeq
We substitute this into the WRG eq.~(\ref{beta=0}) and expand it about $x \approx 0$. 
Then if we choose the coupling constants and the anomalous dimension that satisfy this equation, the theory satisfies $\beta=0$.
\beq
\gamma&=&\frac{N+1}{2 \pi^2} g_2,\\
g_3&=&\frac{2(3N+5)}{3(N+2)}g_2^2 +\frac{2\pi^2}{3(N+2)} g_2,\label{g_3}\\
g_4&=&3g_2 g_3 -\frac{2(N+7)}{3(N+3)}g_2^3 +\frac{\pi^2}{N+3}g_3\nonumber\\
&=&\frac{1}{3(N+2)(N+3)}\Big( (16N^2 +66N+62)g_2^3 + 2\pi^2 (6N+14)g_2^2+2\pi^4 g_2 \Big).\nonumber\\
\label{g_4}
\eeq
Note that all the coupling constants are written in terms of $g_2$ only.
Similarly, we can fix all the coupling constants $g_n$ using $g_2$ order by order.
The function $f(x)$ with such coupling constants describes the conformal field theory and has one free parameter, $g_2$.
In other words, if we fix the value of $g_2$, we obtain a conformal field theory.

For example, we choose 
\beq
g_2&=&-\frac{1}{2} \cdot \frac{2\pi^2}{N+1} \equiv -\frac{1}{2}a.
\eeq
Then the function $f(x)$ become 
\beq
f(x) =\frac{1}{a } \ln (1+ a x ),\label{CP-UV}
\eeq
and this is the K\"{a}hler potential of the C$P^N$ model at UV fixed point.
In fact, the function (\ref{CP-UV}) satisfies the condition (\ref{beta=0}) exactly.

From this discussion, we find that one of the novel ${\bf SU}(N)$ symmetric conformal field theories is identical to the UV fixed point theory of the C$P^N$ model for a specific value of $g_2$.
In this case, the symmetry of this theory is enhanced to ${\bf SU}(N+1)$ because the C$P^N$ model has the isometry ${\bf SU}(N+1)$.

\section{Summary and Discussion}
The nonlinear $\sigma$ model in three dimensions is nonrenormalizable in perturbation theory.
We have studied ${\cal N}=2$ supersymmetric NL$\sigma$Ms using WRG equation, which is one of nonperturbative methods. 

First, we examined the sigma models whose target spaces are the Einstein-K\"{a}hler manifolds.
We have shown that the theories whose target spaces are compact Einstein-K\"{a}hler manifolds with positive scalar curvature have two fixed points.
One of them is the Gaussian IR fixed point and the other is the nontrivial UV fixed point. We can define the continuum limit at this UV fixed point by the fine-tuning of the bare coupling constant. In this sense, NL$\sigma$Ms on Einstein-K\"{a}hler manifolds with positive scalar curvature are renormalizable in three dimensions. At this point, the scaling dimension of all superfields is zero, as in the two dimensional theories. On the other hand, the theories whose target spaces are Einstein-K\"{a}hler manifolds with negative scalar curvature (for example $D^2$ with the Poincar\'{e} metric) have only an Gaussian IR fixed point, and cannot have a continuum limit.

Second, we investigated the phase structures of both $CP^{N-1}$ and $Q^{N-2}$ models using large-$N$ method.
The $CP^{N-1}$ model has two phases: the $SU(N)$ symmetric and broken phases.
In the symmetric phase, all dynamical fields have mass ($m$) due to the vacuum expectation value of the auxiliary field $M$.
The auxiliary fields also have a mass ($2m$), and the field $M$ corresponds to the bound state of two dynamical spinor field $\bar{\psi}^i \psi^i$.
On the other hand, in the broken phase, a dynamical field has a vacuum expectation value, and global $SU(N)$ symmetry is broken.
There are $(N-1)$ massless Nambu-Goldstone bosons and their superpartners, and non of the auxiliary fields are physical particles.

The $Q^{N-2}$ model has three phases: the Chern-Simons, Higgs and $SO(N)$ broken phases.
In the Chern-Simons and Higgs phase, all dynamical fields have a mass ($m$) due to the vacuum expectation value of the auxiliary field $M$ and $A_0$, respectively.
The auxiliary fields also have a mass ($2m$), and the field $M$ corresponds to the bound state of the fermion and anti-fermion; $M \sim \bar{\psi}^i \psi^i$, and $A_0$ corresponds to the bound state of two fermions $A_0 \sim \bar{\psi}^{ci} \psi^i$.
In the Chern-Simons phase, the gauge fields obtain the mass due to the induced Chern-Simons interaction.
In the Higgs phase, the gauge field acquire the mass through the Higgs mechanism and the imaginary part of the field $A_0$ is absorbed by the gauge field.
In the broken phase, two components of dynamical fields  $\varphi^{N-1}$ and $\varphi^N$ have the vacuum expectation values, then global $SO(N)$ symmetry is broken down to $SO(N-2)$.
There are $(N-2)$ massless Nambu-Goldstone bosons and their superpartners, and non of the auxiliary fields are physical particles.

Finally, we constructed a class of the ${\bf SU}(N)$ symmetric conformal field theory using the WRG equation. This class has one free parameter, $g_2$, corresponding to the anomalous dimension of the scalar fields. If we choose a specific value of this parameter, we recover the conformal field theory defined at the UV fixed point of the C$P^N$ model and the symmetry is enhanced to ${\bf SU}(N+1)$.

We argued that a certain class of ${\cal N}=2$ supersymmetric NL$\sigma$Ms are renormalizable in three dimensions. A similar argument may also be valid for ${\cal N}=1$ supersymmetric NL$\sigma$Ms and NL$\sigma$Ms without supersymmetry. Especially, NL$\sigma$Ms for Einstein manifolds with positive scalar curvature might be renormalizable, as long as we use Riemann normal coordinates.


\section*{Acknowledgements}
We would like to thank Tim. Morris, Dmitri Kazakov and Andre Leclair for useful discussions. 
This work was supported in part by the Grant-in-Aid for Scientific
Research (\#16340075 and \#13135215).
E.I is supported by Research Fellowship of the Japan Society for the Promotion of Science (JSPS) for Young Scientists (No.16-07971).





\end{document}